# LLM-based Framework for Bearing Fault Diagnosis

Laifa Tao[1,2,3,4], Haifei Liu[2,3,4], Guoao Ning[2,3,4], Wenyan Cao[2,3,4], Bohao Huang[2,3,4], Chen Lu[1,2,3,4],*

[1]Hangzhou International Innovation Institute, Beihang University, Hangzhou, China
[2]Institute of Reliability Engineering, Beihang University, Beijing, China
[3]Science & Technology on Reliability & Environmental Engineering Laboratory, Beijing, China
[4]School of Reliability and Systems Engineering, Beihang University, Beijing, China
*Corresponding author, luchen@buaa.edu.cn

E-mails Information for Authors: taolaifa@buaa.edu.cn (L.T.); phoebeliu@buaa.edu.cn (H.L.); sy2214210@buaa.edu.cn (G.N.); 20376030@buaa.edu.cn (W.C.); HuangBohao@buaa.edu.cn (B.H.); luchen@buaa.edu.cn (C.L.)

**Abstract**

Accurately diagnosing bearing faults is crucial for maintaining the efficient operation of rotating machinery. However, traditional diagnosis methods face challenges due to the diversification of application environments, including cross-condition adaptability, small-sample learning difficulties, and cross-dataset generalization. These challenges have hindered the effectiveness and limited the application of existing approaches. Large language models (LLMs) offer new possibilities for improving the generalization of diagnosis models. However, the integration of LLMs with traditional diagnosis techniques for optimal generalization remains underexplored. This paper proposed an LLM-based bearing fault diagnosis framework to tackle these challenges. First, a signal feature quantification method was put forward to address the issue of extracting semantic information from vibration data, which integrated time and frequency domain feature extraction based on a statistical analysis framework. This method textualized time-series data, aiming to efficiently learn cross-condition and small-sample common features through concise feature selection. Fine-tuning methods based on LoRA and QLoRA were employed to enhance the generalization capability of LLMs in analyzing vibration data features. In addition, the two innovations (textualizing vibration features and fine-tuning pre-trained models) were validated by single-dataset cross-condition and cross-dataset transfer experiment with complete and limited data. The results demonstrated the ability of the proposed framework to perform three types of generalization tasks simultaneously. Trained cross-dataset models got approximately a 10% improvement in accuracy, proving the adaptability of LLMs to input patterns. Ultimately, the results effectively enhance the generalization capability and fill the research gap in using LLMs for bearing fault diagnosis.

## 1 Introduction

Bearings, as universal components in rotating machinery, will inevitably experience faults during operation. Consequently, bearing fault diagnosis is critical for detecting and isolating faults.

It is fundamental for the health monitoring and lifespan management of rotating machinery, and significantly impacts the reliability and safety of modern industrial systems (Matania et al., 2024). Currently, bearing fault diagnosis faces challenges in generalization in three aspects.

First, the operating environment affects rotating machinery during real-world operations. As technology advances, the operating conditions of bearings have become increasingly complex, resulting in diverse cross-condition data distributions (Su et al., 2022). However, achieving comprehensive condition analysis under actual conditions is challenging. It is necessary to conduct effective and accurate knowledge transfer and fault diagnosis under variable conditions. For example, wind speed, temperature, load, and operation time all influence bearing performance in wind turbines. The operating condition of the bearings varies greatly under high load at low temperatures with high wind speeds versus low load at high temperatures with low wind speeds. Multi-operating conditions caused by complex environments in real-world applications also pose challenges for fault data collection. Therefore, it is crucial to diagnose bearing faults in unknown conditions using data in known conditions (Lu et al., 2024).

Second, due to the high reliability requirements of industrial systems, faults occur randomly and infrequently, resulting in fewer fault samples than normal (Wu et al., 2024a). For instance, commercial aircraft engines, known for their reliability, may record only a few faults over thousands of flight hours, leading to a problem of fault samples for bearing fault diagnosis. This imbalance can cause classification bias, i.e., the models tend to misclassify fault samples as normal (Wang et al., 2024). Achieving high-precision diagnosis with small and imbalanced fault samples remains a widespread research focus (López et al., 2013).

Third, in practical applications, various types of bearings are used in equipment, such as sliding bearings, rolling bearings, deep groove ball bearings, angular contact ball bearings, and self-aligning ball bearings. Even within the same type of bearing, different models from different manufacturers are chosen based on specific installation positions and uses. For example, a fault diagnosis method suitable for rotating machinery in submarines may not apply to aviation machinery, and the applicable technique is generally discussed in each of the two application scenarios. Previous researches often require re-adaptation of feature extraction and classification methods or retraining neural networks for different bearing data, leading to complex operations and weak generalization in practical engineering. Thus, there is an urgent need for intelligent techniques that can quickly adapt to different types of bearings, improving product development and delivery efficiency while reducing repetitive efforts and development costs, thereby enhancing industrial productivity. Therefore, these place a requirement on cross-dataset fault diagnosis. To sum up, bearing fault diagnosis must improve its generalization capability in complex scenarios such as cross-condition, small-sample, and cross-dataset applications.

Current research focuses on two main areas for improving generalization in bearing fault diagnosis:

**Cross-condition and small-sample challenges:** Researchers widely adopt transfer learning methods to achieve knowledge transfer from source conditions to target conditions models in this area include the Multi-Scale Deep Subdomain Adaptation Network (MS-DSACNN) (Fu et al., 2024), Time-Spectrum Domain Adaptation Network (TSDAN) (Ding et al., 2024), and Conditional Weighted Transfer Wasserstein Autoencoder (K. Zhao et al., 2023). Additionally, data augmentation models have been used to bridge the gap between ideal and real-world conditions, such as combining digital twins with Generative Adversarial Networks (GANs) (Z. Li et al., 2024), combining Self-Attention Mechanism and Spectral Normalisation (DCGAN) (Zhong et al., 2023), and combining Auxiliary Classifier GAN and Transformer network (Z. Fu et al., 2024) . Also, meta-learning strategies such as Iterative Resampling Deep Decoupling Domain Adaptation (IRDDDA) (Wu et al., 2024b) and Data Reconstruction Hierarchical Recurrent Meta-Learning (DRHRML) (Wu et al., 2024a) have been adopted to enhance the adaptability of the model. Despite their success in addressing cross-condition and small-sample issues, these methods still heavily depend on data, making them less effective for zero-shot and cold-start fault diagnosis. Moreover, most validations are limited to specific datasets, restricting broad generalization. Future research remains to further explore how to use existing models for fault diagnosis of new samples in zero-start scenarios.

**Cross-dataset challenges:** Representing cross-dataset model features is difficult, especially those brought about by insufficient labeled data and inconsistent data distribution in actual measurements. And the adversarial adaptive networks and domain adaptation methods have successfully generalized diagnosis across bearing types within the same equipment or different devices. For instance, Dynamic Multi-Adversarial Adaptation Network (DMAAN) (Tian et al., 2023) and Deep Convolutional Multi-Adversarial Domain Adaptation (DCMADA) (Wan et al., 2022) models use dynamic adjustment and multi-adversarial strategies for effective knowledge transfer and cross-dataset fault diagnosis. Methods like Multi-Branch Domain Adaptation Network (MBDAN) (G. Wang et al., 2023) and Multi-Scale Attention Mechanism Transfer Model (MSATM) (P. Li, 2024), focus on learning high-quality domain-invariant features to handle data variability and scarcity. However, these methods focus on improving cross-dataset fault diagnosis accuracy and rely on feature-intensive data, so their ability to autonomously analyze discrete, unlabeled, and unstructured data is restricted. They also fail to fully utilize the vast knowledge accumulated across devices or domains, limiting their broad applicability.

In summary, there is still room for improvement in the generalization capability of the current bearing fault diagnosis model in cross-condition, small-sample and cross-dataset scenarios. The emergence of Large Language Models (LLMs) offers potential solutions to these challenges. After LLMs are optimized through large-scale parameter fine-tuning, reinforcement learning, and reward mechanisms, they demonstrate strong capabilities in processing sequence data (W. X. Zhao et al., 2023) (Cao et al., 2024). It aligns well with the input forms required for fault diagnosis (Tao et al., 2024) . For example, time series prediction via LLM has been studied, demonstrating the potential of these models for generalized time series analysis tasks: Studies from New York University and Alibaba explored the application of LLMs to time series prediction and multiple time series tasks, respectively (Gruver et al., 2023) (T. Zhou et al., 2023) . Additionally, research from National Yang Ming Chiao Tung University enhanced LLMs' ability to handle time series data by combining time series patches and time encoding (Chang et al., 2023) . At the methodological level, fine-tuning LLMs can improve their generalization to unseen tasks, adapting them to specific inference tasks in particular fields (Lialin et al., 2023). These advances indicate that LLMs can significantly support fault diagnosis by enhancing model adaptability and efficiency. However, there is limited research on combining LLMs with fault diagnosis models to enhance generalization capabilities. This study aims to leverage the advantages of LLMs to improve fault diagnosis models' performance in cross-condition, small-sample, and cross-dataset scenarios.

This study constructed a fault diagnosis feature system and proposed a vibration data model fine-tuning framework based on LoRA and QLoRA. The fault diagnosis feature system systematically identified and integrated key fault indicators and parameters to build a comprehensive feature system that effectively captures equipment operating status and potential fault signals. The model fine-tuning framework introduced a new LLM fine-tuning strategy based on time series data, optimizing hyperparameters to balance time efficiency and prediction accuracy. Therefore, model performance and adaptability in practical applications are enhanced. Finally, single-dataset, cross-condition, and cross-dataset diagnosis experiments were conducted, comparing complete data transfer and limited data transfer effects. The results showed that LLM-based fault diagnosis methods could enhance the generalization capability of fault diagnosis, meeting cross-condition, small-sample, and cross-dataset practical engineering needs.

The main contributions of this study included:

1. Proposed a bearing fault diagnosis feature system that vectorizes vibration data into text, expanding traditional fault diagnosis methods and enabling integration with LLMs.

2. Introduced an efficient vibration data model fine-tuning framework, validated across bearing datasets through cross-condition, small-sample, and cross-dataset case studies, demonstrating the study's generalized diagnosis capabilities for various bearing fault modes.

The paper is structured as follows: Section 1 introduces the research background and content; Section 2 provides an overview of related work; Section 3 presents the proposed methods based on features and data for LLM-based bearing diagnosis; Section 4 displays the case studies supporting the proposed methods; Section 5 concludes with a summary of findings and innovations.

## 2 Related Work

### 2.1 Bearing fault diagnosis

Bearing fault diagnosis in industrial applications faces challenges such as cross-condition, small-sample, and cross-dataset adaptation. Enhancing the generalization capability of fault diagnosis models is essential. Typically, fault diagnosis relies on analyzing vibration signals in complex environments with multiple conditions. The need for model improvement focuses on three main areas: (1) Retention of long-term knowledge from vibration signals: The ability of long-term intellectual memory allows the model to be more sensitive to the cross-condition and small-sample challenges. Therefore, they can capture minor vibrations, temperature changes, and other key metrics anomalies, and effectively predict potential failures. This feature is particularly important because in cross-condition scenarios, fault performance may vary with the environment and conditions, and small sample problems mean that there is very limited data available for training. Continuous learning from even a few fault cases enhances the model's sensitivity to new or rare fault types, supporting continuous health monitoring and intelligent maintenance strategies (An et al., 2022). As a result, the automation and intelligence of system maintenance strategies are improved to cope with the cross-condition and small-sample challenges. This deep knowledge accumulation and flexible application are keys to ensuring efficient and accurate fault prediction. In this regard, An et al. proposed an LSTM-based fault diagnosis method for rolling bearings, which can efficiently process and memorize long-term dependencies in time-series data by utilizing the structure of a recurrent neural network (RNN) to learn and memorize fault patterns in historical data. (2) Extraction of meta-knowledge from vibration signals: Meta-knowledge is abstracted from multiple data sources or tasks, enabling models to adapt to new or rare environments. Mechanical equipment operates in cross-condition scenarios, often with scarce fault data. Using meta-knowledge reduces the need for retraining, making better use of limited data and reducing costs and time. For instance, L. M. Wang discussed an envelope demodulation analysis of sensitive IMFs to obtain a fault character frequency strategy, which can be applied to abstract more robust meta-knowledge of fault metrics from multiple data sources to enhance cross-condition generalization capability (Y. Zhang & Wang, 2020). J. Li developed an empirical wavelet transform (EWT) to extract intrinsic modulation information through an orthogonal basis by decomposing the signal decomposition into single components on an orthogonal basis to extract the intrinsic modulation information (J. Li et al., 2020). Therefore, a variety of dynamically changing environments can be automatically identified and adapted to using the meta-knowledge obtained from wavelet decomposition, thus enhancing model generalization. H. Li proposed a methodology to reduce the variance of data distributions, helping design algorithms to identify and abstract cross-dataset and cross-condition common features, and improve the adaptability and predictive power of models in new environments (H. Li et al., 2021). (3) Rapid parameter tuning for vibration signal diagnosis models: Reducing time and resources for parameter adjustment improves real-time fault response efficiency. Qian and Qin propose an improved conditional distribution alignment mechanism for effective cross-condition fault diagnosis (Qian et al., 2023).enhancing cross-domain adaptive capabilities. This mechanism allows for effective transfer learning, so that the models can quickly adapt to new environments and fault types. Therefore, the need for retraining and manual adjustments is reduced.

Despite these advancements, traditional fault diagnosis models often struggle with retaining long-term dependencies and complex patterns due to simpler feature learning and fewer parameters. Many models prioritize stability over flexibility. As a result, they adapt slowly to new conditions, so retraining or manual intervention is required. To address these issues, this paper proposed a bearing fault diagnosis framework based on LLMs, integrating LLM's long-term memory capabilities with fine-tuning techniques. This framework leveraged LLMs' strengths in handling complex data patterns, enhancing learning and memory capabilities for various fault modes. By fine-tuning a few parameters, the framework could quickly adapt to new conditions and fault types. Therefore, the limitations of traditional models in flexibility and adaptation speed are overcome.

These improvements can significantly enhance the efficiency and accuracy of bearing fault diagnosis systems, stabilizing the overall operation of rotating machinery.

2.2 Series analysis based on LLM

LLMs possess the capability to store vast amounts of unstructured knowledge using billions of parameters, effectively forming long-term memory (X. Lu et al., 2024) . Leveraging attention mechanisms and neural networks, LLMs can extract crucial information from sequences. It demonstrates initial success in time series analysis applications such as speech, video, and anomaly detection. For instance, Shruthi Hassan Sathish at San Jose State University combined CNN-RNN and Transformer models for multimodal emotion classification in image (video) and sound (audio) sequences (Sathish, 2023) . Mi Zhou and colleagues from Southern Power Grid incorporated meta-learning with LLMs to classify electrical time series data (M. Zhou et al., 2023). At the Belgrade Astronomical Observatory, Evgeny A. Smirnov employed the GPT-4-vision-preview model to classify asteroid vibration states (Smirnov, 2024).

Additionally, LLMs learn representations and sequence relationships from large-scale data through pre-training and generalize to specific applications via fine-tuning, performing well under varied and scarce data conditions (Guo et al., n.d.). Tian Zhou from Alibaba developed an LLM framework for time series prediction, which proved effective in various fields, including power, healthcare, and traffic, and delivered promising results under few-shot conditions (T. Zhou et al., 2023).

Bearing fault diagnosis involves identifying anomalies in vibration time series data, which is a typical time series classification problem. Given the complex operating conditions of bearings, diagnosis methods must perform well in cross-condition and small sample scenarios. Thus, the application of LLMs for bearing fault diagnosis is feasible and promising.

2.3 Supervised fine-tuning

To facilitate the quick application of pre-trained models in specific fields, supervised fine-tuning is commonly employed. This approach simplifies the training process and allows for rapid updates (N. Ding et al., 2023). E. J. Hu et al. (Hu et al., 2021) introduced the LoRA method, which improves training efficiency by keeping part of the layer fixed weights and adding trainable low-rank decomposition matrices. T. Dettmers et al. (Dettmers et al., 2023) proposed the QLoRA supervised fine-tuning method, which reduces memory usage and achieves better results on small, high-quality datasets. Muhammad Najam Dar et al. (Dar et al., 2022) demonstrated cross-dataset learning using pre-trained models and fine-tuning, verifying robustness in practical applications. Supervised fine-tuning can improve the adaptability and accuracy of pre-trained models while reducing training costs, and enable rapid parameter updates and the application of general knowledge from pre-trained models to downstream tasks. Therefore, it achieves good performance in cross-condition, cross-dataset, and small-sample tasks. In bearing fault diagnosis, data collection is costly and fault rarely occurs (J. Zhang et al., 2024), the resulting in scarce fault samples and a small sample size for model training (Chen et al., 2023). This necessitates high model generalization capabilities, making supervised fine-tuning a suitable approach for applying pre-trained models to bearing fault diagnosis. Hongyu Zhong et al. (Zhong et al., 2023) alleviated the need for large amounts of raw data in bearing fault diagnosis by fine-tuning the lower layers of the target network with a small amount of data to adapt to new tasks. Chuanjiang Li et al. (C. Li et al., 2023) designed a scaling and translation fine-tuning strategy to quickly adapt to the changing requirements of fault diagnosis tasks.

2.4 Gap analysis

Facing challenges such as cross-condition, small-sample, and cross-dataset adaptation, this study integrated LLMs with fine-tuning techniques to address the limitations of current technologies, specifically in terms of knowledge retention and adaptability. The goal is to meet the practical requirements for long-term knowledge memory, meta-knowledge extraction, and rapid parameter tuning in fault diagnosis models. Firstly, we developed a fault diagnosis feature system based on vibration data and performed deep decoupling of meta-knowledge within this data. Secondly, given LLMs' strong capabilities in handling sequential data, memory retention, and knowledge generalization, along with their successful supervised fine-tuning applications across various specific fields, we applied LLMs to the field of fault diagnosis. Lastly, a fine-tuning framework was designed to enable rapid parameter tuning of the fault diagnosis LLM model. This framework aims to enhance the efficiency and accuracy of fault diagnosis in cross-condition scenarios.

## 3 Proposed Methods

The proposed framework for bearing fault diagnosis using LLMs is outlined as follows. In the fault diagnosis feature construction based on vibration data, extracting both time-domain and frequency-domain features of vibration signals could create a fine-tuning dataset for the LLMs. Using LoRA and QLoRA methods, the LLM was fine-tuned for fault diagnosis. The fine-tuning framework involved segmenting the vibration signal into patches, converting these into LLM input dimensions using value and token embedding layers. Finally, the LLM was fine-tuned to achieve fault diagnosis.

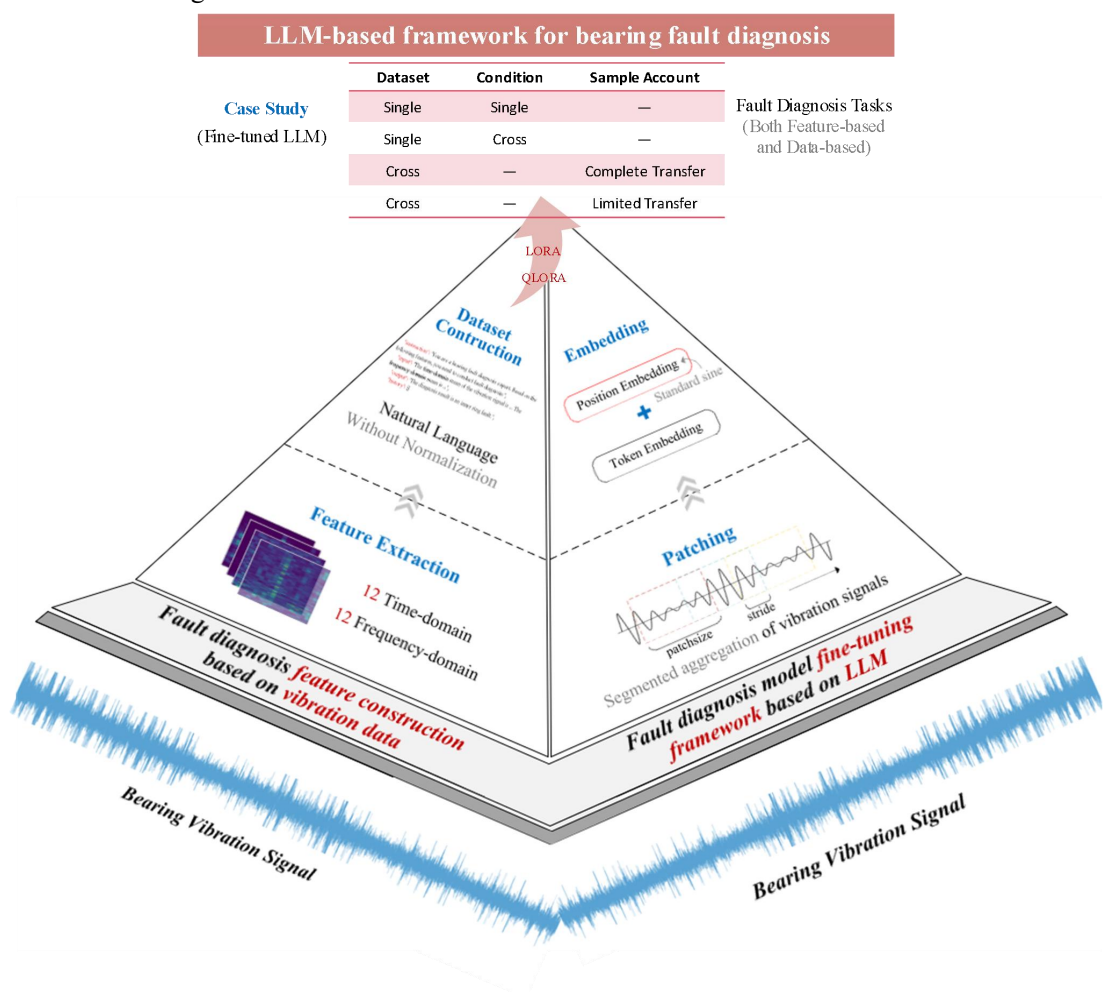

Figure 1 The overall process of the proposed method

3.1 Fault diagnosis feature construction based on vibration data

3.1.1　Feature Extraction

This method utilizes time-domain vibration signals from bearings, which are discrete data points collected by sensors, distinguishing it from traditional LLM frameworks. To utilize the LLM's capability to process semantic information, interpretable information is extracted from the raw vibration signals. A feature-based LLM fault diagnosis method is proposed, extracting both time-domain and frequency-domain features from the raw vibration data. Also, as a framework, the features could also be gained from other methods, taking the permutation entropy-based 2D feature extraction for example (Landauskas et al., 2020).

Commonly used time-domain features include mean value, standard deviation, square root amplitude, absolute mean value, peak value, skewness, kurtosis, variance, kurtosis index, peak index, waveform index, and pulse index. Generally, frequency analysis refers to analyzing a signal following Fourier transform. Commonly used frequency-domain analysis methods include spectrum, power spectrum, and cepstrum. The frequency-domain features of vibration signals commonly used include frequency mean value, frequency variance, frequency skewness, frequency kurtosis, gravity frequency, frequency standard deviation, frequency root mean square, average frequency, regularity degree, variation parameter, eighth-order moment, and sixteenth-order moment.

We selected 12 time-domain and 12 frequency-domain features, totaling 24 features, as shown in Table 1. These are simpler parameters, and complex time-frequency analysis methods like EMD decomposition or wavelet packet analysis were not used for extracting complex or localized features of vibration signals. There are two reasons: First, the extraction of complex features is cumbersome and may need to be designed for the features of vibration data; second, we want to make full use of the learning, extraction, and classification capabilities of the LLM for fault diagnosis based on the easily accessible time-domain and frequency-domain features.

Table 1 Time-domain and frequency-domain features

| Time domain Feature | Formula | Frequency domain Feature | Formula |
|---|---|---|---|
| Mean value | $p_1 = \dfrac{1}{N}\sum_{n=1}^{N} x(n)$ | Frequency mean value | $p_{13} = \dfrac{1}{K}\sum_{k=1}^{K} s(k)$ |
| Standard deviation | $p_2 = \sqrt{\dfrac{1}{N-1}\sum_{n=i}^{N}[x(n)-p_1]^2}$ | Frequency variance | $p_{14} = \dfrac{1}{K-1}\sum_{k=1}^{K}[s(k)-p_{13}]^2$ |
| Square root amplitude | $p_3 = \left(\dfrac{1}{N}\sum_{n=1}^{N}\sqrt{|x(n)|}\right)^2$ | Frequency skewness | $p_{15} = \dfrac{1}{Kp_{14}^{\frac{3}{2}}}\sum_{k=1}^{K}[s(k)-p_{13}]^3$ |
| Absolute mean value | $p_4 = \dfrac{1}{N}\sum_{n=1}^{N}|x(n)|$ | Frequency kurtosis | $p_{16} = \dfrac{1}{Kp_{14}^2}\sum_{k=1}^{K}[s(k)-p_{13}]^4$ |
| Peak value | $p_5 = max|x(n)|$ | Gravity frequency | $p_{17} = \dfrac{\sum_{k=1}^{K} f_k s(k)}{\sum_{k=1}^{K} s(k)}$ |
| Skewness | $p_6 = \dfrac{1}{N}\sum_{n=1}^{N}(x(n))^3$ | Frequency standard deviation | $p_{18} = \sqrt{\dfrac{\sum_{k=1}^{K}(f_k-p_{17})^2 s(k)}{K\sum_{k=1}^{K} s(k)}}$ |
| Kurtosis | $p_7 = \dfrac{1}{N}\sum_{n=1}^{N}(x(n))^4$ | Frequency root mean square | $p_{19} = \sqrt{\dfrac{\sum_{k=1}^{K} f_k^2 s(k)}{\sum_{k=1}^{K} s(k)}}$ |
| Variance | $p_8 = \dfrac{1}{N}\sum_{n=1}^{N}(x(n))^2$ | Average frequency | $p_{20} = \sqrt{\dfrac{\sum_{k=1}^{K} f_k^4 s(k)}{\sum_{k=1}^{K} f_k^2 s(k)}}$ |
| Kurtosis index | $p_9 = p_7/(\sqrt{p_6})^2$ | Regularity degree | $p_{21} = \dfrac{\sum_{k=1}^{K} f_k^2 s(k)}{\sqrt{\sum_{k=1}^{K} s(k) \sum_{k=1}^{K} f_k^4 s(k)}}$ |
| Peak index | $p_{10} = max\|x(n)\|/p_2$ | Variation parameter | $p_{22} = p_{18}/p_{17}$ |
| Waveform index | $p_{11} = p_2/p_4$ | Eighth-order moment | $p_{23} = \dfrac{\sum_{k=1}^{K}(f_k - p_{17})^3 s(k)}{Kp_{18}^3}$ |
| Pulse index | $p_{12} = max\|x(n)\|/p_4$ | Sixteenth-order moment | $p_{24} = \dfrac{\sum_{k=1}^{K}(f_k - p_{17})^4 s(k)}{Kp_{18}^4}$ |

### 3.1.2  Fine-tuning dataset construction and fine-tuning

After extracting time-domain and frequency-domain features, we converted them into a format that the LLM can understand by using language descriptions instead of direct numerical values. As shown in Figure 2, we combine the values of the time-frequency domain features extracted above with their textual descriptions, while using the fault modes as labels for supervised learning, with *inputs* in the form of question-and-answer pairs, with the aim that the model is able to give the correct outputs when inputting instructions and input in new tasks. This approach avoided normalizing the features, preserving their physical significance.

It is worth mentioning that in order to make the model rely as little as possible on expert knowledge, our linguistic descriptions of the textual parts, except for the features, are only used as an illustrative example. The main purpose is to enable the LLM to understand the task to be performed and the meaning of the input data.

> "instruction": "You are a bearing fault diagnosis expert. Based on the following features, you need to conduct fault diagnosis:",
> "input": "The **time-domain** mean of the vibration signal is ... The **frequency-domain** mean is ...",
> "output": "The diagnosis result is an inner ring fault.",
> "history": []

Figure 2 Example of textualized model input

Fine-tuning helps the LLM learn the relationships and knowledge associated with specific feature parameters. As shown in Figure 3, the LoRA and QLoRA are employed to fine-tuning methods, allowing the LLM to learn patterns within the fine-tuning data for fault diagnosis classification.

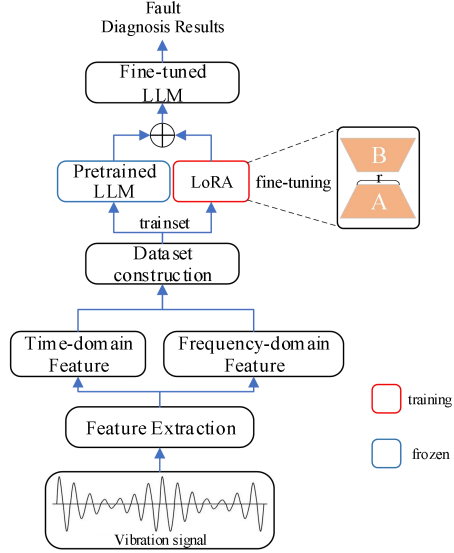

Figure 3 Framework of feature-based LLM fault diagnosis

### 3.2 Fault diagnosis model fine-tuning framework based on LLM

The LLM-based fault diagnosis model is illustrated below. After instance normalization of the vibration data, we segmented the signal into patches and converted each patch into an LLM input dimension vector using value and position embeddings. Then these vectors were input into the LLM. We extracted features by fine-tuning the Add & Layer Norm layers of LLM and classified the data by the Layer normalization and Linear layers.

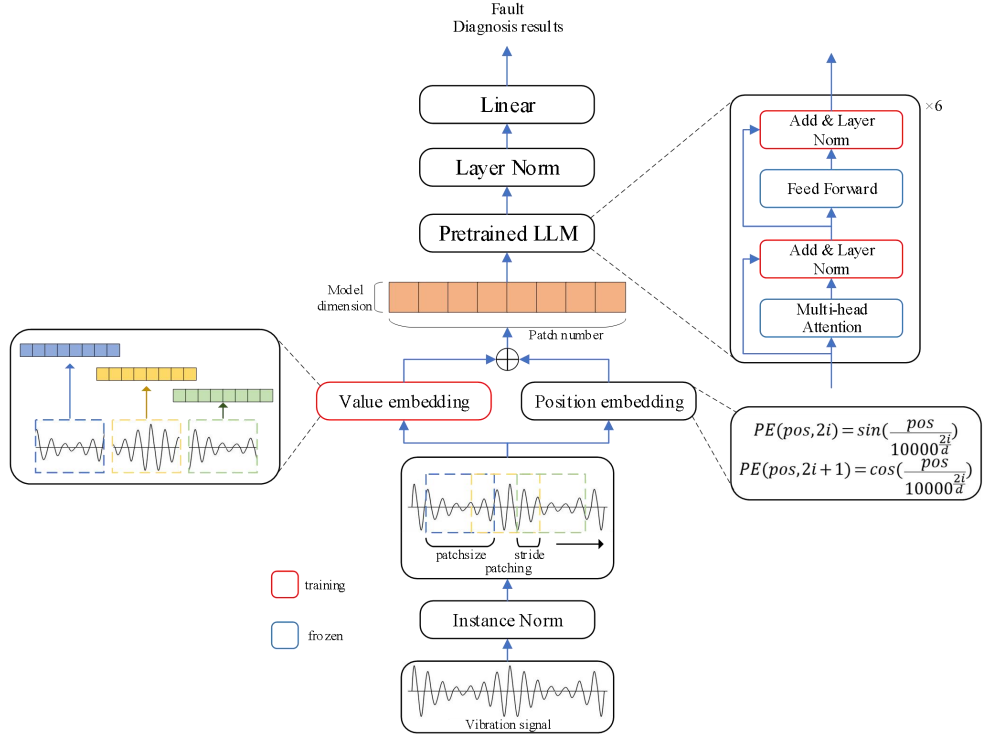

Figure 4 Framework of data-based LLM fault diagnosis

#### 3.2.1 Patching

The high sampling frequency of vibration data creates information redundancy. To address this, we use a patching method from PatchTST to aggregate adjacent data points. This method extracted local semantic information to enable the LLM to focus on aggregated patches' features and patterns, reduce redundancy and improve local information processing. Patching also reduced token count in LLMs, increasing input data length for more information and reducing computational load and memory usage.

3.2.2 Embedding

Token embedding: Each token in the input text was converted into a vector of the LLM input dimension through token embedding. To input vibration data into the LLM, we also needed to perform token embedding on the vibration data, converting it into the LLM input dimension. Due to the patching of the bearing vibration data, we used a one-dimensional convolutional layer to convert each merged patch into the LLM input dimension.

Position embedding: For position embedding, we used the standard sinusoidal position embedding of the transformer to embed the position information directly into the representation of the sequence. The sinusoidal position embedding formula is shown below.

$$\vec{p_t}^{(i)} = f(t)^{(i)} := \begin{cases} \sin(\omega_k t), & \text{if } i = 2k \\ \cos(\omega_k t), & \text{if } i = 2k+1 \end{cases}$$

$$\omega_t = \frac{1}{10000^{2k/d}}$$

Therefore, following token embedding and position embedding on the bearing vibration data, the embedded values were added up as the final input embedding value to be delivered to the LLM.

3.2.3 Freeze attention layer

In our GPT-2 model, we froze the multi-head attention and FFN layers, and only trained the layer norm and position embedding layers. The multi-head attention and FFN layers of the model contained most of the knowledge learned by the model in the pre-training phase. Therefore, this module was frozen in order to: 1) make full use of the knowledge learned by the model to help realize the fault diagnosis of bearing vibration data, and 2) to reduce the number of trainable parameters and accelerate the training process since it contained the vast majority of the model's parameters.

3.2.4 Normalization

We instance normalize the input vibration data using mean and variance while adding affine transformations to the LayerNorm layer of the gpt2 model by training its learnable affine transformation parameters to further facilitate knowledge transfer. The learnable affine transformation adjusts the offset and scaling of the data by learning the parameters while normalizing to increase the expressive power of the model thus allowing the model to learn more complex data distributions and patterns. At the same time, through affine changes, the model can better adapt to different data distributions and tasks, and improve the model's generalization capability, thus enabling it to have better task performance on different datasets.

The affine transformation formula is shown below:

$$y = \frac{x - E[x]}{\sqrt{Var[x] + \varepsilon}} * \gamma + \beta$$

Where $\gamma$ and $\beta$ are learnable affine transformation parameters and $\varepsilon$ is a very small value to avoid a denominator of 0 in case the variance is 0.

3.2.5 Loss function

The cross-entropy loss function was used to measure the difference between predicted and actual labels:

$$L_{CCE} = -\frac{1}{N}\sum_{i=1}^{N}\sum_{c=1}^{C} y_{i,c} \cdot log\ (p_{i,c})$$

**4 Case Study**

Based on the above proposed LLM-based bearing fault diagnosis framework, validation experiments were carried out for feature-based LLM fault diagnosis and data-based LLM, respectively. It was proved that the proposed framework could cross-condition, small-sample, and cross-dataset diagnose faults through the single-dataset experiments, the single-dataset cross-condition experiments, the complete-data and limited-data cross-dataset experiments at the same time.

Table 2 Content and validated capabilities for case study

| Section | Dataset | Condition | Sample Account | Diagnosis Capability |
|---|---|---|---|---|
| 4.4.1 | Single | Single | — | Simple scenario |
| 4.5.2 | | | | |
| 4.4.2 | Single | Cross | — | Cross working conditions |
| 4.5.3 | | | | |
| 4.4.3.1 | Cross | — | Complete data Transfer | Cross datasets |
| 4.5.4.1 | | | | |
| 4.4.3.2 | Cross | — | Limited data Transfer | Small sample Cross datasets |
| 4.5.4.2 | | | | |

4.1 Dataset introduction

In the case study, experiments were conducted on feature-based LLM fault diagnosis and data-based LLM fault diagnosis using four public bearing fault diagnosis datasets: CWRU, MFPT, JNU, and PU.

In the CWRU dataset, we used fault data from the drive-end bearing with a sampling rate of 12 kHz. The faults had depths of 0.007 inches, 0.014 inches, and 0.021 inches, covering four operating conditions with different loads and speeds. There were four fault modes: normal, inner race fault, outer race fault, and rolling element fault.

In the MFPT dataset, data from three normal bearings, 3+7 outer race fault bearings, and seven inner race fault bearings were used. This dataset included three outer race fault bearings had an input shaft speed of 25 Hz, a load of 270 lbs, and a sampling frequency of 97,656 Hz. The seven outer race fault bearings had an input shaft speed of 25Hz, a sampling frequency of 48,828 Hz, and loads of 25, 50, 100, 150, 200, 250, and 300 lbs. The seven inner race fault bearings had an input shaft speed of 25 Hz, a sampling frequency of 48,828 Hz, and loads of 0, 50, 100, 150, 200, 250, and 300 lbs.

The JNU dataset had a sampling frequency of 50 kHz and included three rotational speeds: 600 rpm, 800 rpm, and 1000 rpm. It covered four fault modes: normal, inner race fault, outer race fault, and rolling element fault.

In the PU dataset, we used data from 12 artificially damaged bearings (seven with outer race faults and five with inner race faults) and six normal bearings. The vibration signal sampling frequency was 64 kHz, with three fault modes.

The STFT time-frequency images of the data used in the case study showed significant differences in vibration data across the four datasets as below. Studying their generalization capabilities is of research significance.

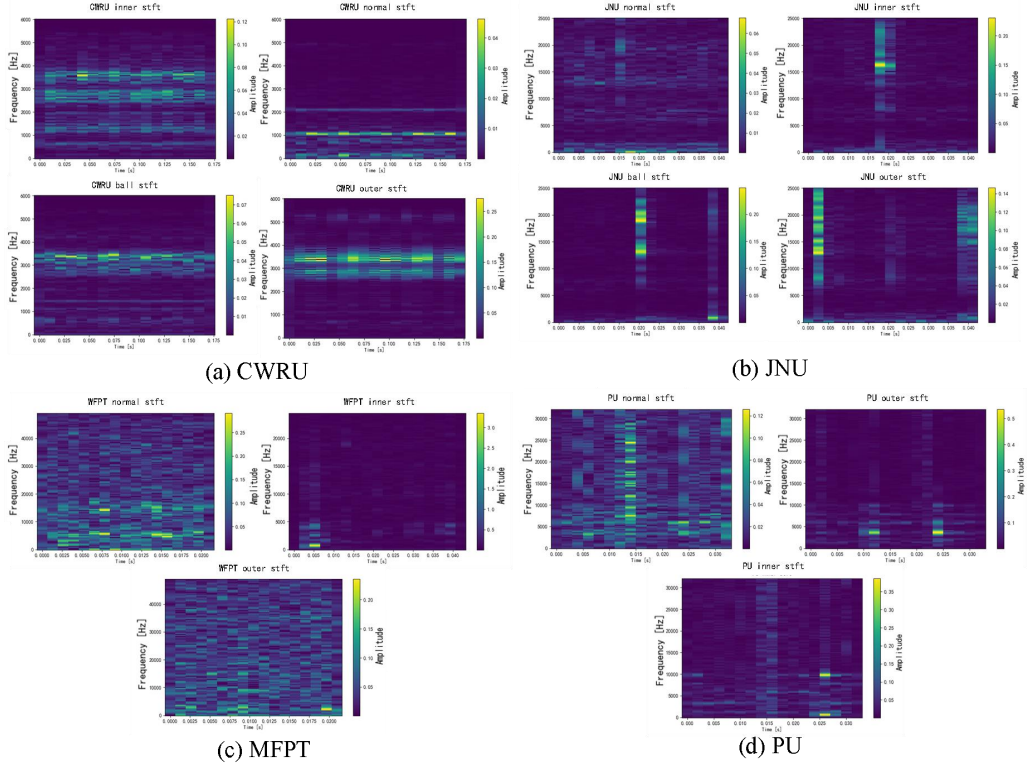

(a) CWRU  (b) JNU
(c) MFPT  (d) PU

Figure 5 STFT images of four datasets for different failure modes

4.2 Data preprocessing

The bearing vibration data are divided into samples of 2048 points using a sliding window method, with a step size smaller than the window length to retain potential fault information at the window edges and minimize information loss.

For feature-based LLM fault diagnosis experiments, time-domain and frequency-domain features were extracted from these samples using the formulas mentioned in the second section to build the fine-tuning dataset. The training and testing sets were split 8:2. For data-based LLM fault diagnosis experiments, the training, validation, and testing sets were split 8:1:1. To leverage the LLM's capability to learn from large datasets, we ensured that each fault mode in each dataset had an equal number of samples, aiming to minimize the impact of class imbalance and to maximize the knowledge learned from the data.

For subsequent experiments on model generalization with combined and cross-training on different datasets, we classified each bearing dataset by fault mode (normal, inner race fault, outer race fault, and rolling element fault) without further label distinction based on operating conditions or fault severity due to the possibly great differences in these cross-dataset factors.

4.3 Experimental environment

The experimental setup is as follows: Nvidia GeForce RTX 4090 GPU with 24GB memory; development tools were VScode and torch 2.1.0; CUDA version 12.2.

4.4 Feature-based LLM fault diagnosis case study

In the feature-based LLM bearing fault diagnosis case study, we used the ChatGLM2-6B-chat model, an open-source bilingual dialogue model developed by THUDM of Tsinghua University. The model's weights are fully open for academic research, and its bilingual training facilitates fine-tuning with Chinese data for fault diagnosis. Due to the high stability requirements of fault diagnosis tasks, we set the temperature coefficient to 0.01 to ensure consistent results from the LLM.

4.4.1 Single-dataset experiment

The diagnosis results for a single dataset are shown in the table below. The results indicated that the LLM effectively recognized and identified the features of bearing vibration signals for fault diagnosis, and its accuracy improved as training epochs increased. It can be seen that its fault diagnosis capabilities were robust.

Table 2 Results of feature-based single-dataset experiments

| Epoch | CWRU | MFPT | JNU | PU |
|---|---|---|---|---|
| | Accuracy | | | |
| 1 | 0.5919 | 0.8294 | 0.6945 | 0.5020 |
| 2 | 0.6935 | 0.9644 | 0.8325 | 0.6984 |
| 3 | 0.8109 | 0.9697 | 0.8904 | 0.7599 |
| 4 | 0.9072 | 0.9844 | 0.9046 | 0.7857 |
| 5 | 0.9405 | 0.9917 | 0.9148 | 0.8224 |
| 6 | 0.9527 | 0.9936 | 0.9182 | 0.8274 |
| 7 | 0.9597 | 0.9963 | 0.9165 | 0.8254 |
| 8 | 0.9615 | 0.9963 | 0.9182 | 0.8333 |
| 9 | 0.9632 | 0.9954 | 0.9165 | 0.8353 |
| 10 | 0.9685 | 0.9982 | 0.9182 | 0.8403 |

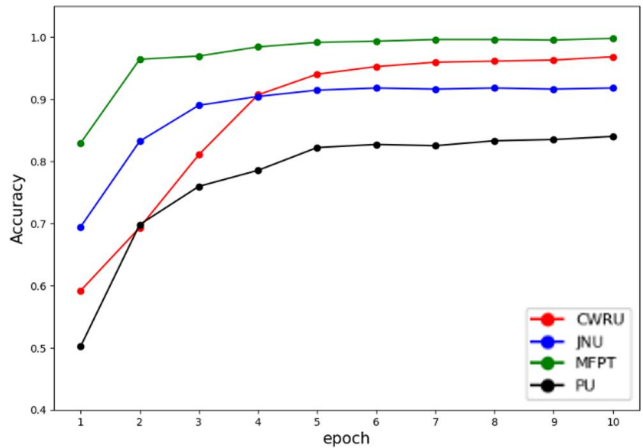

Figure 6 Results of feature-based single-dataset experiments

4.4.2 Single-dataset cross-condition experiment

To assess the generalization capability of the proposed method under unseen conditions, cross-condition diagnosis experiments were conducted using the CWRU dataset. The results for 10 training epochs are shown in the table. The feature-based LLM fault diagnosis method was able to diagnose cross-condition faults, with higher accuracy when diagnosing within the same condition extrapolation than inter-condition extrapolation. More condition data also improved the diagnosis accuracy for new conditions.

Table 3 Working condition of the CWRU dataset

| Working Condition No. | Approx. Motor Speed (rpm) | Motor Load (HP) |
|---|---|---|
| 0 | 1797 | 0 |
| 1 | 1772 | 1 |
| 2 | 1750 | 2 |
| 3 | 1730 | 3 |

Table 4 Cross-condition experiment results of the CWRU dataset

| No. | Train Condition | Test Condition | Accuracy (Epoch=10) |
|---|---|---|---|
| 1 | 0,1 | 2 | 0.7074 |
| 2 | 0,1 | 3 | 0.6769 |
| 3 | 1,2 | 0 | 0.6419 |
| 4 | 1,2 | 3 | 0.7293 |
| 5 | 0,2 | 1 | 0.7686 |
| 6 | 0,2 | 3 | 0.6463 |
| 7 | 1,3 | 2 | 0.8559 |
| 8 | 1,3 | 0 | 0.6332 |
| 9 | 0,3 | 1 | 0.7074 |
| 10 | 0,3 | 2 | 0.6638 |
| 11 | 2,3 | 0 | 0.6245 |
| 12 | 2,3 | 1 | 0.7948 |
| 13 | 1,2,3 | 0 | 0.6550 |
| 14 | 0,2,3 | 1 | 0.8646 |
| 15 | 0,1,2 | 3 | 0.7249 |
| 16 | 0,1,3 | 2 | 0.8996 |

4.4.3 Cross-dataset experiment

4.4.3.1 Complete data transfer experiment

Cross-dataset experiments were conducted using the CWRU, MFPT, JNU, and PU datasets. The model was trained on three of these datasets and tested it on the fourth. The design and results of these experiments are presented in the table below. Compared with single dataset training, multi-dataset training improved the LLM's knowledge transfer to new datasets. For the four datasets, the diagnosis effect was greatly improved compared with single-dataset training at transfer training under epoch=1. For the PU and CWRU datasets, the diagnosis accuracy was improved by 0.077 and 0.0235, respectively, after 10 training epochs with the other three datasets. In contrast, improvements for the MFPT and JNU datasets were smaller.

Table 5 Complete-data cross-dataset experiment results

| Train Dataset (epoch=10) | CWRU+MFPT +JNU | CWRU+MFPT +PU | CWRU+JNU +PU | MFPT+JNU +PU |
|---|---|---|---|---|
| Transfer dataset | PU | JNU | MFPT | CWRU |
| Test dataset | PU | JNU | MFPT | CWRU |
| Epoch | Accuracy | | | |
| 1 | 0.8607 | 0.8688 | 0.9918 | 0.9391 |
| 2 | 0.8738 | 0.9063 | 0.9973 | 0.9760 |
| 3 | 0.8929 | 0.9120 | 0.9991 | 0.9840 |
| 4 | 0.8935 | 0.9182 | 1.0000 | 0.9872 |
| 5 | 0.8976 | 0.9114 | 0.9991 | 0.9808 |
| 6 | 0.9024 | 0.9211 | 1.0000 | 0.9840 |
| 7 | 0.9101 | 0.9182 | 0.9991 | 0.9872 |
| 8 | 0.9131 | 0.9177 | 0.9991 | 0.9904 |
| 9 | 0.9113 | 0.9211 | 1.0000 | 0.9888 |
| 10 | 0.9173 | 0.9194 | 1.0000 | 0.9920 |

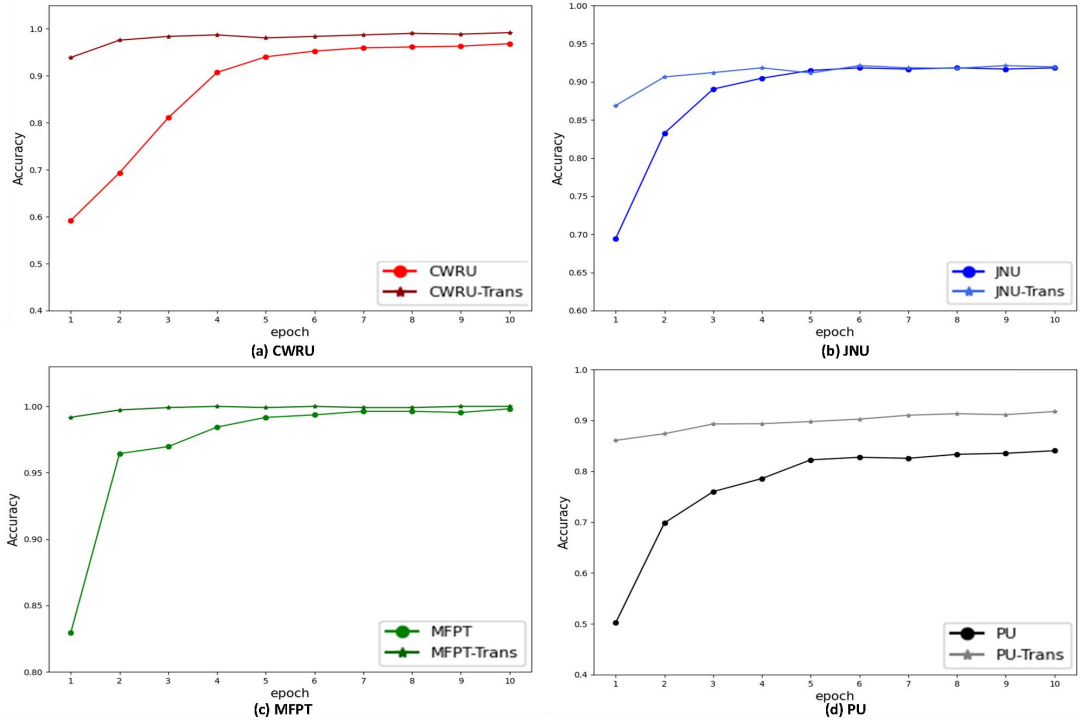

Figure 7 Complete-data cross-dataset experiment results

4.4.3.2 Limited data transfer experiment

Considering the actual needs of engineering applications and conditions of use, less failure data may be available for new equipment. To simulate practical engineering conditions with limited fault data, we fine-tuned the model with 10% of the new dataset and compared the results with those trained only on 10% of the new dataset on the basis of the original training on multiple datasets to validate the model's impact on the auxiliary diagnosis performance of new small sample data based on old knowledge. The experimental design and results are shown in the table below.

Table 6 Limited data cross-dataset experiment results

| Train Dataset (epoch=10) | CWRU+MFPT +JNU | CWRU+MFPT +PU | CWRU+JNU +PU | MFPT+JNU +PU |
|---|---|---|---|---|
| Transfer dataset | 10% PU | 10% JNU | 10% MFPT | 10% CWRU |
| Test dataset | PU | JNU | MFPT | CWRU |
| Epoch | Accuracy | | | |
| 1 | 0.8899 | 0.6616 | 0.6396 | 0.4423 |
| 2 | 0.8982 | 0.6945 | 0.6268 | 0.6314 |
| 3 | 0.8756 | 0.6775 | 0.6442 | 0.6987 |
| 4 | 0.8935 | 0.6934 | 0.6405 | 0.7660 |
| 5 | 0.8833 | 0.6996 | 0.6469 | 0.8093 |
| 6 | 0.8946 | 0.7036 | 0.6460 | 0.8125 |
| 7 | 0.8845 | 0.6746 | 0.6487 | 0.8349 |
| 8 | 0.8768 | 0.6672 | 0.6487 | 0.8109 |
| 9 | 0.8720 | 0.6672 | 0.6460 | 0.8029 |
| 10 | 0.8690 | 0.6746 | 0.6460 | 0.8253 |

Table 7 Limited data experiment results

| Train Dataset | 10% PU | 10% JNU | 10% MFPT | 10% CWRU |
|---|---|---|---|---|
| Test dataset | PU | JNU | MFPT | CWRU |

| Epoch | Accuracy | | | |
|---|---|---|---|---|
| 1 | 0.1304 | 0.4991 | 0.0502 | 0.0000 |
| 2 | 0.4155 | 0.5060 | 0.5119 | 0.0000 |
| 3 | 0.4560 | 0.5645 | 0.5301 | 0.0000 |
| 4 | 0.4637 | 0.6371 | 0.6058 | 0.2308 |
| 5 | 0.4446 | 0.6366 | 0.6852 | 0.5064 |
| 6 | 0.4369 | 0.6440 | 0.6223 | 0.2548 |
| 7 | 0.4601 | 0.6457 | 0.6487 | 0.3157 |
| 8 | 0.5202 | 0.6536 | 0.5985 | 0.2484 |
| 9 | 0.5482 | 0.6587 | 0.6122 | 0.3189 |
| 10 | 0.5815 | 0.6740 | 0.6387 | 0.3157 |

4.5 Data-based LLM fault diagnosis case study

4.5.1 Hyperparameter selection experiment

To determine the optimal hyperparameters for bearing fault diagnosis, we optimized the patch size and stride based on the JNU dataset. The experimental design and results are shown in the table below. The results showed that a larger patch size and smaller stride parameter achieved higher diagnosis accuracy. However, due to longer training times with a smaller stride, we selected a patch size of 128 and a stride of 8 for a balance of accuracy and training time.

Table 8 Results of hyperparameter selection experiment

| Patch Size | Stride | Accuracy |
|---|---|---|
| 256 | 8 | $0.9943^{+0.0032}_{-0.0032}$ |
| 128 | 64 | $0.9853^{+0.0024}_{-0.0024}$ |
| 128 | 32 | $0.9936^{+0.0032}_{-0.0032}$ |
| 128 | 16 | $0.9928^{+0.0032}_{-0.0032}$ |
| 128 | 8 | $0.9955^{+0.0024}_{-0.0024}$ |
| 128 | 4 | $0.9970^{+0.0019}_{-0.0019}$ |
| 64 | 32 | $0.9807^{+0.0061}_{-0.0061}$ |
| 64 | 16 | $0.9894^{+0.0010}_{-0.0010}$ |
| 64 | 8 | $0.9932^{+0.0016}_{-0.0016}$ |
| 64 | 4 | $0.9981^{+0.0011}_{-0.0011}$ |
| 32 | 16 | $0.9796^{+0.0025}_{-0.0025}$ |
| 32 | 8 | $0.9909^{+0.0016}_{-0.0016}$ |
| 32 | 4 | $0.9955^{+0.0012}_{-0.0012}$ |
| 16 | 8 | $0.9803^{+0.0060}_{-0.0060}$ |
| 16 | 4 | $0.9841^{+0.0065}_{-0.0065}$ |
| 8 | 4 | $0.9758^{+0.0030}_{-0.0030}$ |

4.5.2 Single-dataset experiment

The main hyperparameters for the single dataset experiment were patch size=128, stride=8, learning rate=0.001, and epoch=50. After each training epoch, the model was validated using the validation set, and the best-performing model was selected as the final one. The experimental results are shown in the table below.

Table 9 Results of single-dataset experiment

| Dataset | Accuracy |
|---|---|
| CWRU | $0.9996^{+0.0004}_{-0.0006}$ |
| MFPT | $0.9982^{+0.0015}_{-0.0015}$ |
| JNU | $0.9981^{+0.0006}_{-0.0006}$ |
| PU | $0.9239^{+0.0051}_{-0.0051}$ |

4.5.3 Single-dataset cross-condition Experiment

In order to verify the generalizability of the proposed method for unseen working conditions,

a single-dataset cross-condition fault diagnosis experiment was also designed for the data-based LLM fault diagnosis method. The specific experimental design and results are shown in the table below. The CWRU dataset specific operating conditions are described in Section 4.1.

From the experimental results, it could be seen that the proposed method had good diagnosis performance for single-dataset cross-case diagnosis. Specifically, the working condition inference case had a higher diagnosis accuracy than the extrapolation case, while a larger amount of working condition data also had a certain improvement effect on the diagnosis accuracy. It suggested that LLM can learn the distribution of failure modes under different operating conditions and apply them to new operating conditions to realize cross-condition fault diagnosis.

Table 10 Cross-condition experiment results of CWRU dataset

| No. | Train Condition | Test Condition | Accuracy |
|---|---|---|---|
| 1 | 0,2 | 1 | $0.9976^{+0.0024}_{-0.0024}$ |
| 2 | 0,2 | 3 | $0.9888^{+0.0065}_{-0.0065}$ |
| 3 | 1,3 | 2 | $0.9963^{+0.0037}_{-0.0064}$ |
| 4 | 1,3 | 0 | $0.9832^{+0.0103}_{-0.0103}$ |
| 5 | 0,3 | 1 | $0.9753^{+0.0204}_{-0.0204}$ |
| 6 | 0,3 | 2 | $0.9607^{+0.0232}_{-0.0232}$ |
| 7 | 0,1 | 2 | $0.9900^{+0.0079}_{-0.0079}$ |
| 8 | 0,1 | 3 | $0.9570^{+0.0210}_{-0.0210}$ |
| 9 | 1,2 | 0 | $0.9707^{+0.0267}_{-0.0267}$ |
| 10 | 1,2 | 3 | $0.9951^{+0.0035}_{-0.0035}$ |
| 11 | 2,3 | 1 | $0.9853^{+0.0147}_{-0.0153}$ |
| 12 | 2,3 | 0 | $0.9667^{+0.0223}_{-0.0223}$ |
| 13 | 0,1,2 | 3 | $0.9814^{+0.0022}_{-0.0022}$ |
| 14 | 1,2,3 | 0 | $0.9901^{+0.0072}_{-0.0072}$ |
| 15 | 0,1,3 | 2 | $0.9967^{+0.0033}_{-0.0046}$ |
| 16 | 0,2,3 | 1 | $1.0000^{+0.0000}_{-0.0000}$ |

Cross-case experiments were also conducted on the JNU bearing dataset. The JNU dataset consists of three conditions, which are described in the following table:

Table 11 Working condition of JNU dataset

| Working Condition No. | Approx. Motor Speed (rpm) |
|---|---|
| 1 | 600 |
| 2 | 800 |
| 3 | 1000 |

The specific experimental design and results are shown in the table below:

Table 12 Cross-condition experiment results of JNU dataset

| No. | Train Condition | Test Condition | Accuracy |
|---|---|---|---|
| 1 | 1,3 | 2 | $0.9796^{+0.0064}_{-0.0064}$ |
| 2 | 1,2 | 3 | $0.9439^{+0.0227}_{-0.0227}$ |
| 3 | 2,3 | 1 | $0.8991^{+0.0363}_{-0.0363}$ |

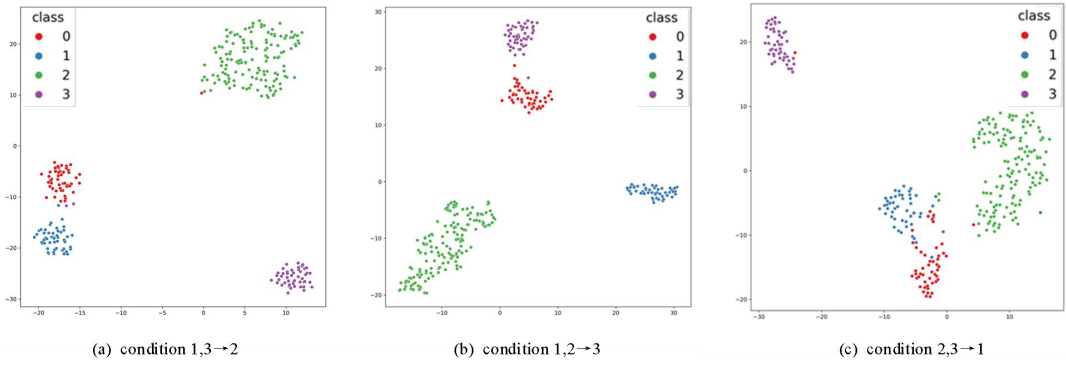

(a) condition 1,3→2  (b) condition 1,2→3  (c) condition 2,3→1

Figure 8 Feature visualization based on t-SNE of JNU cross-condition experiment

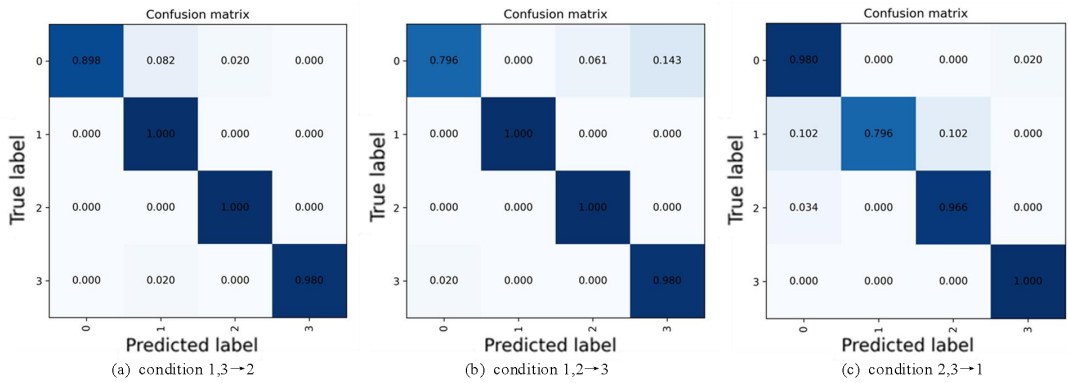

(a) condition 1,3→2  (b) condition 1,2→3  (c) condition 2,3→1

Figure 9 Test confusion matrix of JNU cross-condition experiment
(Label ball inner normal outer)

The specific working conditions of the PU dataset, and the PU data cross-condition fault diagnosis results are shown in the tables below.

Table 13 Working condition of PU dataset

| Working Condition No. | Rotational Speed (rpm) | Load Torque (Nm) | Radial Force (N) |
| --- | --- | --- | --- |
| 0 | 1500 | 0.7 | 1000 |
| 1 | 900 | 0.7 | 1000 |
| 2 | 1500 | 0.1 | 1000 |
| 3 | 1500 | 0.7 | 400 |

Table 14 Cross-condition experiment results of PU dataset

| No. | Train Condition | Test Condition | Accuracy |
| --- | --- | --- | --- |
| 1 | 0,2 | 1 | $0.4852^{+0.0478}_{-0.0478}$ |
| 2 | 0,2 | 3 | $0.7722^{+0.0060}_{-0.0060}$ |
| 3 | 1,3 | 2 | $0.6704^{+0.0238}_{-0.0238}$ |
| 4 | 1,3 | 0 | $0.6500^{+0.0363}_{-0.0363}$ |
| 5 | 0,3 | 1 | $0.5889^{+0.0520}_{-0.0520}$ |
| 6 | 0,3 | 2 | $0.9232^{+0.0079}_{-0.0079}$ |
| 7 | 0,1 | 2 | $0.8537^{+0.0107}_{-0.0107}$ |
| 8 | 0,1 | 3 | $0.6982^{+0.0261}_{-0.0261}$ |
| 9 | 1,2 | 0 | $0.8908^{+0.0053}_{-0.0053}$ |

| 10 | 1,2 | 3 | $0.7232^{+0.0112}_{-0.0112}$ |
| 11 | 2,3 | 1 | $0.4852^{+0.0138}_{-0.0138}$ |
| 12 | 2,3 | 0 | $0.9259^{+0.0102}_{-0.0102}$ |
| 13 | 0,1,2 | 3 | $0.7398^{+0.0203}_{-0.0203}$ |
| 14 | 1,2,3 | 0 | $0.8973^{+0.0045}_{-0.0045}$ |
| 15 | 0,1,3 | 2 | $0.8658^{+0.0094}_{-0.0094}$ |
| 16 | 0,2,3 | 1 | $0.5648^{+0.0167}_{-0.0167}$ |

#### 4.5.4 Cross-dataset experiment

4.5.4.1 Complete-data transfer experiment

To fully utilize the LLM's capabilities and verify the impact of multi-dataset training on the generalization of LLM-based diagnosis, we designed a multi-dataset training diagnosis experiment. The experimental design and results are shown in the table below. The comparison with single-dataset complete-data experiments is illustrated in Figure 10. For the PU dataset, multi-dataset training significantly improved diagnosis accuracy by approximately 1%. For the CWRU, MFPT, and JNU datasets, the impact of multi-dataset training was minimal.

Table 15 Complete-data cross-dataset experiment results

| No. | Train Dataset | Continued Training Dataset | Test Dataset | Accuracy |
|---|---|---|---|---|
| 1 | CWRU+MFPT+JNU | PU | PU | $0.9338^{+0.0096}_{-0.0096}$ |
| 2 | CWRU+MFPT+PU | JNU | JNU | $0.9958^{+0.0011}_{-0.0011}$ |
| 3 | CWRU+JNU+PU | MFPT | MFPT | $0.9988^{+0.0008}_{-0.0008}$ |
| 4 | MFPT+JNU+PU | CWRU | CWRU | $0.9987^{+0.0010}_{-0.0010}$ |

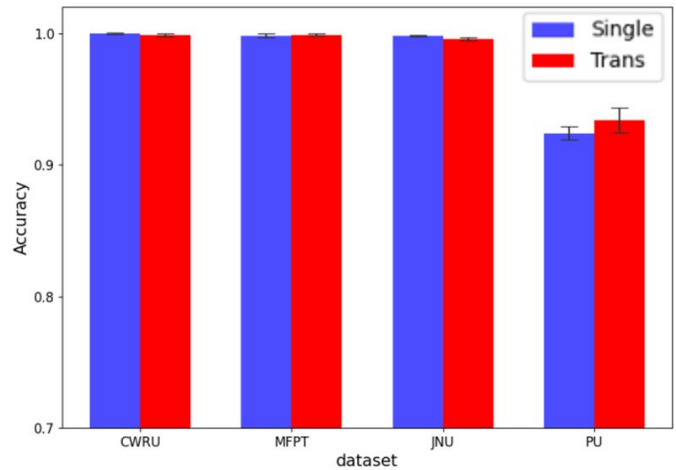

Figure 10 The comparison with single-dataset complete-data experiments

4.5.4.2 Limited data transfer experiment

A limited data transfer experiment was conducted for the data-based LLM fault diagnosis method. Following training on multiple datasets, the model underwent fine-tuning with 10% of a new dataset. The results were compared with those from a scenario where the model was trained solely on 10% of the new dataset. The test dataset was the full test set rather than a subset of the small data. Multiple trials were conducted simultaneously, and the average diagnosis accuracy was taken to reduce experimental error. The experimental design and results are shown in the table below. By comparing the experimental results with Figure 11, it was found that the multi-dataset training improved accuracy and reduced the standard deviation for small data scenarios, suggesting that LLMs can transfer the learned knowledge to new datasets and improve the generalization and stability of the model on new datasets. The most significant improvements were observed for the PU, MFPT, and JNU datasets, with accuracy increases by 0.0396, 0.0376, and 0.0327, respectively, while the CWRU dataset saw a minimal increase of 0.0028.

Table 16 Few data cross-dataset experiment results

| No. | Train Dataset | Transfer Dataset | Test Dataset | Accuracy |
|---|---|---|---|---|
| 1 | CWRU+MFPT+JNU | 10% PU | PU | $0.7872^{+0.0084}_{-0.0084}$ |
| 2 | CWRU+JNU+PU | 10% MFPT | MFPT | $0.9495^{+0.0075}_{-0.0075}$ |
| 3 | CWRU+MFPT+PU | 10% JNU | JNU | $0.8644^{+0.0238}_{-0.0238}$ |
| 4 | MFPT+JNU+PU | 10% CWRU | CWRU | $0.9634^{+0.0037}_{-0.0037}$ |
| 5 | 10% PU | / | PU | $0.7476^{+0.0153}_{-0.0153}$ |
| 6 | 10% MFPT | / | MFPT | $0.9118^{+0.0275}_{-0.0275}$ |
| 7 | 10% JNU | / | JNU | $0.8317^{+0.0230}_{-0.0230}$ |
| 8 | 10% CWRU | / | CWRU | $0.9606^{+0.0157}_{-0.0157}$ |

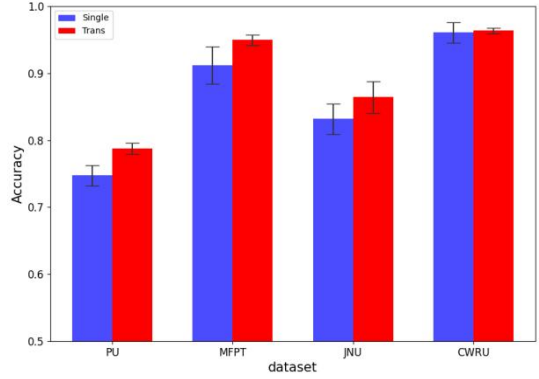

Figure 11 Limited data cross-dataset experiment results

## 5 Conclusion

To address the challenges of cross-condition adaptability, small-sample learning difficulties, and cross-dataset generalization in bearing fault diagnosis, this paper integrated the long-term memory characteristics of LLMs with fine-tuning techniques. We proposed an LLM-based bearing fault diagnosis framework that included a signal feature quantification method combining time-domain and frequency-domain feature extraction, as well as vibration data fine-tuning methods based on LoRA and QLoRA. This method effectively improved fault diagnosis accuracy and enhances the model's generalization capability. The case studies demonstrated the adaptability of the proposed framework across three generalization tasks. Therefore, approximately a 10% improvement was achieved in accuracy after cross-dataset learning.

The innovative contributions of this paper include: It presented a fundamental framework for conducting various tasks in the specific field of fault diagnosis using LLMs. It demonstrated the superior performance of the proposed framework in generalization tasks, and filled the research gap in applying LLMs for bearing fault diagnosis. Initial explorations were made and some results were achieved, but there is still plenty of room for improvement, such as the conducted experiments represent partially generalized scenarios and the discussion of alternative approaches to the feature quantization is not developed. Future research can be conducted on this fundamental framework, such as designing model structures specific to different generalization tasks and using real-world diagnosis tasks to explore the framework's effectiveness, accuracy, and generalization capabilities.

# 6  Declaration

**Funding**: This study was supported by the National Key R&D Program of China under Grant STI 2030—Major Projects (Grant No. 2021ZD0201300), the National Natural Science Foundation of China (Grant Nos. 52472442 and 72471013), and the Research Start-up Funds of Hangzhou International Innovation Institute of Beihang University (Grant Nos. 2024KQ069, 2024KQ036 and 2024KQ035).

**Author Contributions:** Data curation, G.N. and C.L.; Formal analysis, L.T.; Methodology, L.T. and H.L.; Writing—original draft, H.L. and G.N.; Validation, W.C.; Investigation, W.C. and B.H.; Supervision, L.T. and C.L. All authors have read and agreed to the published version of the manuscript.